\documentclass[sigconf]{acmart}

\renewcommand\footnotetextcopyrightpermission[1]{}
\AtBeginDocument{%
  }
\acmConference[Author pre-print]{Accepted at ICSE-SEET '26}{ICSE 2026}{SEET track}

\usepackage{multirow}
\usepackage{graphicx}
\usepackage[most]{tcolorbox}
\usepackage{booktabs}
\usepackage{tabularx}
\usepackage{array}

\begin{document}

\title{AI-Assisted Code Review as a Scaffold for Code Quality and Self-Regulated Learning: An Experience Report}
\author{Eduardo Oliveira}
\orcid{0000-0001-5063-8860}
\email{eduardo.oliveira@unimelb.edu.au}
\affiliation{%
  \institution{The University of Melbourne}
  \city{Melbourne}
  \country{Australia}}

\author{Michael Fu}
\orcid{0000-0001-7211-3491}
\email{michael.fu@unimelb.edu.au}
\affiliation{%
  \institution{The University of Melbourne}
  \city{Melbourne}
  \country{Australia}}

\author{Patanamon Thongtanunam}
\orcid{0000-0001-6328-8839}
\email{patanamon.t@unimelb.edu.au}
\affiliation{%
  \institution{The University of Melbourne}
  \city{Melbourne}
  \country{Australia}}
  
\author{Sonsoles López-Pernas}
\email{sonsoles.lopez@uef.fi}
\orcid{0000-0002-9621-1392}
\affiliation{%
  \institution{University of Eastern Finland}
  \city{Joensuu}
  \country{Finland}
}

\author{Mohammed Saqr}
\orcid{0000-0001-5881-3109}
\email{mohammed.saqr@uef.fi}
\affiliation{%
  \institution{University of Eastern Finland}
  \city{Joensuu}
  \country{Finland}
}


\renewcommand{\shortauthors}{Oliveira et al.}

\newcommand{\rqone}{What is student engagement with AI-powered code review in a project-based software engineering subject?}

\newcommand{\rqtwo}{How do students perceive the use of AI-powered code review in a project-based software engineering subject?}

\newcommand{\rqthree}{How does the introduction of AI-powered code review tools reshape or disrupt established code review roles and practices among student teams?}

\newcommand\eduardo[1]{{\textcolor{orange}{Eduardo: #1}}}
\newcommand\pick[1]{{\textcolor{blue}{Pick: #1}}}
\definecolor{mygreen}{rgb}{0.0, 0.5, 0.0}
\newcommand\mike[1]{{\textcolor{mygreen}{Mike: #1}}}

\newcommand{\smallsection}[1]{\noindent\underline{\textbf{#1.}} }

\newtcolorbox{rqanswer}[1][]{
    colback=gray!10,      
    colframe=black,     
    boxrule=0.8pt,          
    #1
}


\begin{abstract}
Code review is central to software engineering education but hard to scale in capstones due to tight deadlines, uneven peer feedback, and limited prior experience. We investigate an LLM-as-reviewer integrated directly into GitHub pull requests (human-in-the-loop) across two cohorts (>100 students, 2023–2024). Using a mixed-methods design—GitHub data, reflective reports, and a targeted survey—we examine engagement and responsiveness as behavioural indicators of self-regulated learning processes. Quantitatively, the 2024 cohort produced more iterative activity (1176 vs. 581 PRs), while technical issues observed in 2023 (227 failed AI attempts) dropped to zero after tool and instructional refinements. Despite different adoption levels (teams using the tool: 93\% vs. 50\%), responsiveness was stable: 32\% (2023) and 33\% (2024) of successfully AI-reviewed PRs were followed by subsequent commits on the same PR. Qualitatively, students used the LLM’s structured comments to focus reviews and discuss code quality, while guidance reduced over-reliance. We contribute: (i) an in-workflow design for an AI reviewer that scaffolds learning while mitigating cognitive offloading; (ii) a repeated cross-sectional comparison across two cohorts in authentic settings; (iii) a mixed-methods analysis combining objective GitHub metrics with student self-reports; and (iv) evidence-based pedagogical recommendations for responsible, student-led AI-assisted review.
\end{abstract}





\maketitle

\section{Introduction}
\label{sec:introduction}

The proliferation of Large Language Models (LLMs) has catalysed a paradigm shift in software engineering practice. LLMs are no longer novelties but are rapidly integrating into professional development workflows, redefining tasks from code generation and debugging to documentation \cite{chen2021evaluating,fan2023large}. This rapid integration presents a critical juncture for software engineering education. As educators, we are tasked with preparing students for an industry where collaboration with AI is becoming the norm. The central question is no longer whether to incorporate these tools, but how to do so in a manner that is pedagogically sound and enhances, rather than hinders, student learning.

One of the most critical industry practices taught in software engineering curricula is code review. In professional settings, it is the foundation for quality assurance, knowledge sharing, and team collaboration \cite{bacchelli2013, sadowski2018modern, patel2024enhancing}. However, translating this practice into an academic setting, particularly in high-stakes software engineering projects, is fraught with challenges. Students often lack the experience to provide insightful feedback, feel uncomfortable criticising their peers, and struggle with the time-consuming nature of manual reviews amidst pressing project deadlines \cite{oliveira2023ai, baresi2025students,parra2025towards, song2020using}. Consequently, peer code reviews in the classroom can be inconsistent, superficial, or neglected altogether, failing to impart the intended skills.

To address these challenges, we present LLM-Reviewer, a custom tool we developed and integrated as an additional code reviewer within the existing pull-request workflow of a Master’s-level software engineering subject. Our tool is implemented as a repository-local automation (GitHub Action), allowing students to invoke it on-demand. Rather than introducing a new platform, it emits standard PR comments alongside human feedback. This design preserved the teams’ branching model, CI/CD pipeline, and sprint cadence, ensuring minimal disruption while capturing all interactions as first-class artefacts in the repository. The study spans two cohorts (2023–2024) and more than 100 students working with real industry clients on authentic projects. Our focus is model-agnostic: we examine the pedagogical and behavioural effects of positioning an LLM inside the established review loop, not the idiosyncrasies of any particular product. We frame our intervention not merely as an automation tool, but as a pedagogical scaffold for fostering Self-Regulated Learning (SRL). This LLM-as-reviewer complements conventional linters and static checkers by providing structured, rubric-aligned natural-language feedback (e.g., rationale, documentation, and design considerations) while keeping humans in the loop for judgement and decisions. By providing immediate, structured feedback designed to prompt reflection---rather than supplying definitive answers---our LLM-Reviewer tool aimed to help students better monitor their work, evaluate its quality, and reflect on their coding practices, thereby fostering the metacognitive skills essential for lifelong learning.

To investigate the impact of our LLM-Reviewer tool for SRL and critical thinking, we use a mixed-methods analysis of repository data, surveys, and detailed student reflections. This study examines student engagement with the tool, their perceptions of its utility, and the broader challenges of its integration. In doing so, this work makes four key contributions to software engineering education: (i) A novel design for an AI review tool that scaffolds learning while mitigating the risk of cognitive offloading; (ii) a detailed repeated cross-sectional comparison between two cohorts on the same subject with a consistent method on the integration of our LLM-Reviewer tool into authentic software engineering settings, offering insights into replicability and stability of our findings; (iii) a mixed methods analysis of student engagement, combining objective behavioral metrics with subjective self-reports and survey responses; and (iv) a set of evidence-based pedagogical recommendations for educators seeking to navigate the new landscape of AI-assisted software engineering.
\section{Background \& Related Work}
\label{sec:background}

\subsection{Code review in education}

In modern software engineering, the pull request (PR) serves as the central mechanism for collaborative development. It is a formal proposal to merge a set of changes---such as new features, bug fixes, or improvements---from a development branch into a main code base~\cite{parra2025towards}. The core of the PR process is code review, a practice widely acknowledged as a basis of quality assurance in professional settings. Industry-based studies have long established that code review is critical for detecting defects, enforcing team-specific coding standards, sharing knowledge across a team, and improving the long-term quality and maintainability of software projects~\cite{bacchelli2013, oliveira2023ai}.

Recognising its industrial importance, software engineering educators have increasingly integrated code review and PR workflows into their curricula, particularly in project-based and software-based capstone courses~\cite{parra2025towards}. The pedagogical goal extends beyond mere defect detection; it is to provide students with authentic experience in critical analysis, collaborative development, and the socio-technical skill of giving and receiving constructive feedback~\cite{oliveira2023ai}. However, translating this practice effectively into the classroom is challenging~\cite{lopez2025dynamics, lopez2025ai}. Studies consistently report that students often lack the technical expertise to provide insightful feedback, leading to reviews that can be superficial or inconsistent~\cite{baresi2025students, bacchelli2013}. Furthermore, significant socio-cognitive barriers exist; students frequently feel uncomfortable critiquing their peers' work due to concerns about fairness, bias, or social dynamics, a contrast to the structured expectations of a professional environment~\cite{oliveira2023ai}. Compounded by the time-consuming nature of manual reviews amidst the pressure of project deadlines, the result is that peer code review in academic settings is often neglected or fails to deliver on its pedagogical promise, creating a clear need for alternative or supplementary approaches.

\subsection{AI-driven feedback as a scaffold for fostering SRL}

The recent emergence of LLMs has introduced a powerful new category of tools into the software engineering landscape. General-purpose LLMs and specialised tools such as GitHub Copilot are now capable of a wide range of development tasks, including code generation, explanation, and debugging~\cite{baresi2025students, parra2025towards}. This has led to the development of AI tools and agents specifically designed to assist with and automate parts of the code review process, aiming to streamline PR approval and enhance developer productivity~\cite{parra2025towards}.

From an educational perspective, the most compelling application of these tools is not as a replacement for human effort, but as a scaffold for student learning \cite{cukurova2025interplay, lopez2025dynamics, misiejuk2025facets}. Their potential can be effectively framed through the lens of SRL, a robust pedagogical theory that describes how individuals actively manage their own learning processes \cite{de2025development}. SRL is often characterised by a cyclical process of planning (setting goals), monitoring (tracking progress and understanding), and evaluating (reflecting on outcomes to inform future actions)~\cite{zimmerman2002becoming}. The pedagogical value of such tools is amplified within the agile methodologies used in project-based courses~\cite{marnewick2023student}, as agile sprints provide a natural framework for students to practice SRL~\cite{linden2018scrum}. However, a critical gap exists: the `evaluate` and `reflect` phases of the student-led agile cycle often lack a consistent and objective feedback mechanism.

Herein lies a key opportunity. An AI-powered review tool can serve as this missing feedback mechanism, acting as a reliable mechanism for the evaluation step~\cite{winne2000measuring}. Unlike peer feedback, which can be delayed, socially complex, and varied in quality, AI-driven feedback is immediate, while its form can be standardised~\cite{zhan2025generative, taranto2020sustaining}. By embedding this feedback directly into the pull request---a space students already own and manage---an intervention can trigger the invisible metacognitive process of self-evaluation visible, tangible and actionable. The goal is not to provide ``answers'' or fix code, but to act as a mirror, inviting students to reflect on their own coding practices. This approach grants them full agency over the timing and nature of their evaluations and responses, empowering them to align their reflective practice with their personal and project goals on their own terms.

While this potential to support SRL is clear, much of the existing work in teaching and promoting code review among software engineering students has focused on their perceptions of LLM usefulness, finding that they value these tools for low-level tasks like coding and often use them as ``learning objects'' to replace sources like Stack Overflow~\cite{baresi2025students, oliveira2023ai}. Current studies rely primarily on self-reported data, leaving a gap in our understanding of students' objective, observable online behaviours \cite{patel2024enhancing, song2025investigating}. Furthermore, while the effectiveness of AI feedback is often discussed, it is rarely linked to tangible outcomes or theoretical educational frameworks like SRL; there is a lack of evidence tracing AI-generated suggestions to subsequent, concrete changes in students' code repositories. Finally, studies often overlook the real-world operational challenges---or ``friction''---that students face when attempting to integrate these tools into the complex workflow of a high-stakes capstone project. Therefore, critical gaps remain in understanding students' actual engagement behaviours, the causal link between AI feedback and code improvement, and the practical challenges of deploying these tools to scaffold learning in authentic educational environments.

\section{Current study}
In this context, we examine how our LLM-Reviewer tool, intentionally designed to foster student agency by making feedback on code quality visible and actionable, can scaffold the evaluation process while mitigating the risk of over-reliance on AI. We report on the integration of our tool in a Master's level software-based capstone, conducted across two semester-long cohorts (2023 and 2024). Building on our prior formative work that established the viability of this tool in a direct comparison with traditional peer review \cite{oliveira2023ai}, this paper shifts its focus to characterise the cross-sectional patterns of student engagement and perception across these two independent cohorts. 

Although LLM tools are rapidly entering professional workflows, empirical evidence in authentic software engineering educational settings remains sparse. Prior work has emphasised perceived usefulness over behaviour traces, and has rarely linked AI feedback to subsequent code changes in repositories. Little is known about the operational friction students face when using AI-assisted review in classrooms, nor about how such tools can scaffold SRL processes of monitoring and evaluation at scale. Addressing these gaps is important for educators who must decide not only whether to adopt AI review in their teaching, but how to deploy it so that it strengthens learning rather than replacing it. Guided by this gap, we investigate the following research questions:
\begin{itemize}
    \item \textbf{RQ1:} How do students engage with an AI-powered code review tool in a capstone project setting?
    \item \textbf{RQ2:} What are students' perceptions of the AI utility for improving code quality and software engineering skills?
\end{itemize}
Answering these questions provides evidence on if and when students use AI review in their projects, whether feedback is acted upon in code, how students experience its benefits and limits, and which failure modes impede adoption. The contribution is twofold. First, we provide a cross-sectional classroom-embedded characterisation of engagement and impact using repository-level traces rather than self-report alone. Second, we introduce scalable and consistent operational measures aligned to SRL that provide a reusable, evidence-based model for designing interventions that foster student agency by making the evaluation process accessible.
\section{Study design}
\label{sec:methods}

We analyse two independent cohorts enrolled in the same subject in 2023 and 2024, examining the consistency of adoption patterns, technical challenges, and responsiveness to AI feedback across cohorts. This study, therefore, employs a mixed-methods approach, using objective repository data and subjective student perceptions to analyse how students engage with the LLM-Reviewer tool as a scaffold for SRL in an authentic capstone setting across two consecutive years.

\subsection{Context and participants}
The study took place in the \textit{Software Project} capstone subject at the University of Melbourne across two offerings: Semester~1, 2023 and Semester~1, 2024. This is a compulsory subject for students in the final year of their Master of Information Technology degree. The course is designed as an authentic, project-based learning experience where students work in teams of five to design, implement, test, and deploy a software product for a real-world industry partner.

A core pedagogical element of the subject is its use of an agile development methodology, with the 14-week semester structured into four multi-week sprints (Weeks 1-4: Design; Weeks 5-8: Development I; Weeks 9-12: Development II; Weeks 13-14: Handover). Each sprint requires teams to plan a set of features, implement them, and demonstrate their work. All teams used GitHub for version control, issue tracking, and managing pull requests, emulating an industrial development environment. This agile, cyclical structure of planning, doing, and reviewing provided a natural framework to support and observe students' SRL processes.

From a cohort of 358 enrolled students (170 in 2023, 188 in 2024), 80 students in 2023 and 26 students in 2024 volunteered to participate and provided informed consent under the university's ethics protocol (Approval \#24272). While non-participants also used the LLM-Reviewer tool as part of the standard curriculum, their data was rigorously excluded from this analysis. This two-cohort design provides a valuable opportunity to validate our findings and observe patterns of adoption in separate, independent groups.

\subsection{AI-powered code review intervention}
Our intervention was designed to scaffold the monitoring and evaluation phases of the SRL cycle by integrating our LLM-Reviewer tool directly into the students' agile development workflow. We used a custom GitHub Action that automated code reviews using OpenAI's GPT-3.5 Turbo model (2023) and OpenAI's GPT-4 model (2024).

In Week~7, towards the end of the first development sprint, students were formally onboarded through a lecture on software quality and code review practices. This included a live demonstration of both traditional peer review and the LLM-Reviewer tool. To motivate engagement, the assessment for Sprints 2 (Week 8) and 3 (Week 12) included two marks (2 out of 15, in each sprint) allocated to the quality of code reviews and subsequent reflections. Our intervention was designed to fit seamlessly into the sprint workflow: at the completion of a development task, (i) a student opens a pull request; (ii) this action automatically triggers our tool, which sends the code changes to the GPT API using the structured prompt detailed in the subsection below; and (iii) the AI's feedback is posted as a comment on the pull request, providing an immediate opportunity for evaluation and reflection before merging.

\underline{Prompt Structure.} A critical component of our intervention was the design of the prompt sent to the LLM. Rather than an open-ended instruction like ``review this code'' or ``fix this code'', we engineered a structured, checklist-based prompt (Figure~\ref{fig:prompt}) based on the code review classification from M{\"a}ntyl{\"a} and Lassenius~\cite{mantyla2008types}. This design was foundational to the tool's pedagogical effectiveness, providing structure, explainability, and a scaffold for learning the dimensions of a high-quality code review.
Our checklist-based design was a deliberate pedagogical choice aimed at preventing cognitive offloading and promoting active student engagement. By providing structured critiques and high-level recommendations without offering ready-to-use code solutions, the intervention is designed to ensure the student remains the central agent in the learning process. It places the cognitive load of understanding the feedback, planning a fix, and implementing it firmly back onto the student. This approach positions the AI as a collaborator that augments a student's ability to evaluate their own work, rather than as a tool that replaces their critical judgment. This ensures a ``human-in-the-loop'' learning process, strengthening the 'evaluate' and 'reflect' phases of SRL instead of allowing students to bypass them \cite{cukurova2025interplay,de2025development,lopez2025llms}.

\begin{figure}[h!]
\begin{verbatim}
Please evaluate the (code) below:
Use the following checklist to guide your analysis:
1. Documentation Defects:
   a. Naming: Assess the quality of software 
      element names.
   b. Comment: Analyse the quality and accuracy of
      code comments.
2. Visual Representation Defects:
   a. Bracket Usage: Identify any issues with
      incorrect or missing brackets.
   b. Indentation: Check for incorrect indentation
      that affects readability.
   c. Long Line: Point out any long code statements
      that hinder readability.
3. Structure Defects:
   a. Dead Code: Find any code statements that
      serve no meaningful purpose.
   b. Duplication: Identify duplicate code statements
      that can be refactored.
4. New Functionality:
   a. Use Standard Method: Determine if a
      standardised approach should be used for
      single-purpose code statements.
5. Resource Defects:
   a. Variable Initialisation: Identify variables
      that are uninitialised or incorrectly
      initialised.
   b. Memory Management: Evaluate the program's
      memory usage and management.
6. Check Defects:
   a. Check User Input: Analyse the validity of
      user input and its handling.
7. Interface Defects:
   a. Parameter: Detect incorrect or missing
      parameters when calling functions or
      libraries.
8. Logic Defects:
   a. Compute: Identify incorrect logic during
      system execution.
   b. Performance: Evaluate the efficiency of the
      algorithm used.

Provide your feedback in a numbered list for each
category.
\end{verbatim}
\caption{The structured, checklist-based prompt used to guide the LLM's code review process.}
\label{fig:prompt}
\end{figure}

\subsection{Data collection and processing}
To observe a holistic view of student engagement using our LLM-Reviewer tool, we computed 5 metrics from GitHub's repository data, which were then systematically processed and aligned.

\paragraph{Data sources:}
\begin{enumerate}
    \item \textbf{Pull Request Metrics:} Metadata for every pull request, including creation timestamps and repository information.
    \item \textbf{Pull Request Comments:} The full text of all comments, including author, type, and timestamp.
    \item \textbf{Pull Request Commits:} All commits associated with a pull request, including their message and timestamp.
    \item \textbf{End-of-Subject Survey:} A voluntary survey with Likert-scale items and one free-text question, mapped to the phases of SRL.
    \item \textbf{Individual Reflective Reports:} Written reflections of approximately 400 words from each student on their experience.
\end{enumerate}

\paragraph{Data processing:}
\begin{itemize}
    \item \textbf{Calendar alignment:} All UTC timestamps were converted to the local timezone (Australia/Melbourne) and mapped to a strict 14-week semester calendar for each respective year. For the 2023 cohort, Week 1 began on Monday, 27 March 2023, and for the 2024 cohort, it began on Monday, 25~March~2024.
    \item \textbf{AI Engagement Classification:} We derived engagement status from the comments data. A pull request was classified as a \textit{Failed AI Attempt (or friction} when the AI bot was triggered but returned an error (for example missing or unauthorised API credentials, non-code artefacts such as \texttt{.png} or \texttt{.zip}, or exceeding file limits). It was a \textit{Successful AI Review} if it contained at least one valid review from a bot user (\verb|cr-gpt[bot]| or \verb|github-actions[bot]|) and the comment matched the tool's template (contained the header string ``ChatGPT review for ...''. All other PRs were classified as \textit{No AI Attempt}. 
    \item \textbf{Action rate metric:} For each PR with a successful AI review, we scanned its associated commits. If at least one commit occurred after the timestamp of the first successful bot comment, the pull request was marked as ``actioned''.
\end{itemize}
This raw data was processed to derive analysis-ready variables.

\subsection{Analysis plan}
Our analysis was structured to directly address our two research questions, with each question drawing from specific data sources to produce key outcomes, and is mirrored in the structure of our Findings and Discussion sections to ensure a clear and consistent narrative.

For \textbf{RQ1}, which investigates \textit{how} students engage, we performed a descriptive statistical analysis on the processed GitHub data. We calculated weekly counts of PRs by their engagement status to understand adoption and technical friction. We then computed a weekly Action Rate---the proportion of successfully reviewed PRs that were followed by a commit---to measure responsiveness. These metrics were visualised in a two-panel figure to show both the composition and the impact of engagement over time.

For \textbf{RQ2}, which examines students’ \textit{perceptions of utility}, we focused on self-reflection reports and survey data. The primary dataset consisted of 400-word reflection reports, worth 8\% of the subject grade in 2023, which yielded a complete set of 80 submissions. These reports captured students’ experiences across the entire code review process. To supplement this, 16 students completed a voluntary survey that provided quantitative context through Likert-scale responses. Following recommended practices for qualitative analysis \cite{nowell2017thematic}, we conducted a thematic analysis of the reflection reports. The first and second authors independently performed initial coding and iteratively refined the codebook by comparing passages with emerging codes. They then collaborated through multiple online meetings to resolve discrepancies and reach a consensus. A final meeting with all three authors was held to review, challenge, and agree on the final themes. We present these themes with exemplar quotes, and where relevant, we integrate supporting evidence from the survey responses, which we mapped to phases of SRL.
\section{Results}
\label{sec:findings}

In this section, we present the findings from our mixed-methods analysis, structured by our research questions. We begin by addressing student engagement through a quantitative analysis of objective data from the teams' GitHub repositories.

\subsection{RQ1: How students engage with the AI review tool}
\smallsection{Approach}
To answer our first research question, we analysed objective behavioural data from all pull requests created by participants in both 2023 and 2024 cohorts. Our analysis reveals a clear narrative of high student intent to use the tool, some technical and pedagogical friction that was largely resolved in the second year, and a consistent pattern of responsiveness to AI feedback across both cohorts. These phenomena are situated within the rhythm of the subject's agile sprints, demonstrating how the AI tool was integrated into an authentic project-based learning workflow.

\begin{figure*}[htbp]
    \centering
    \includegraphics[width=0.6\textwidth]{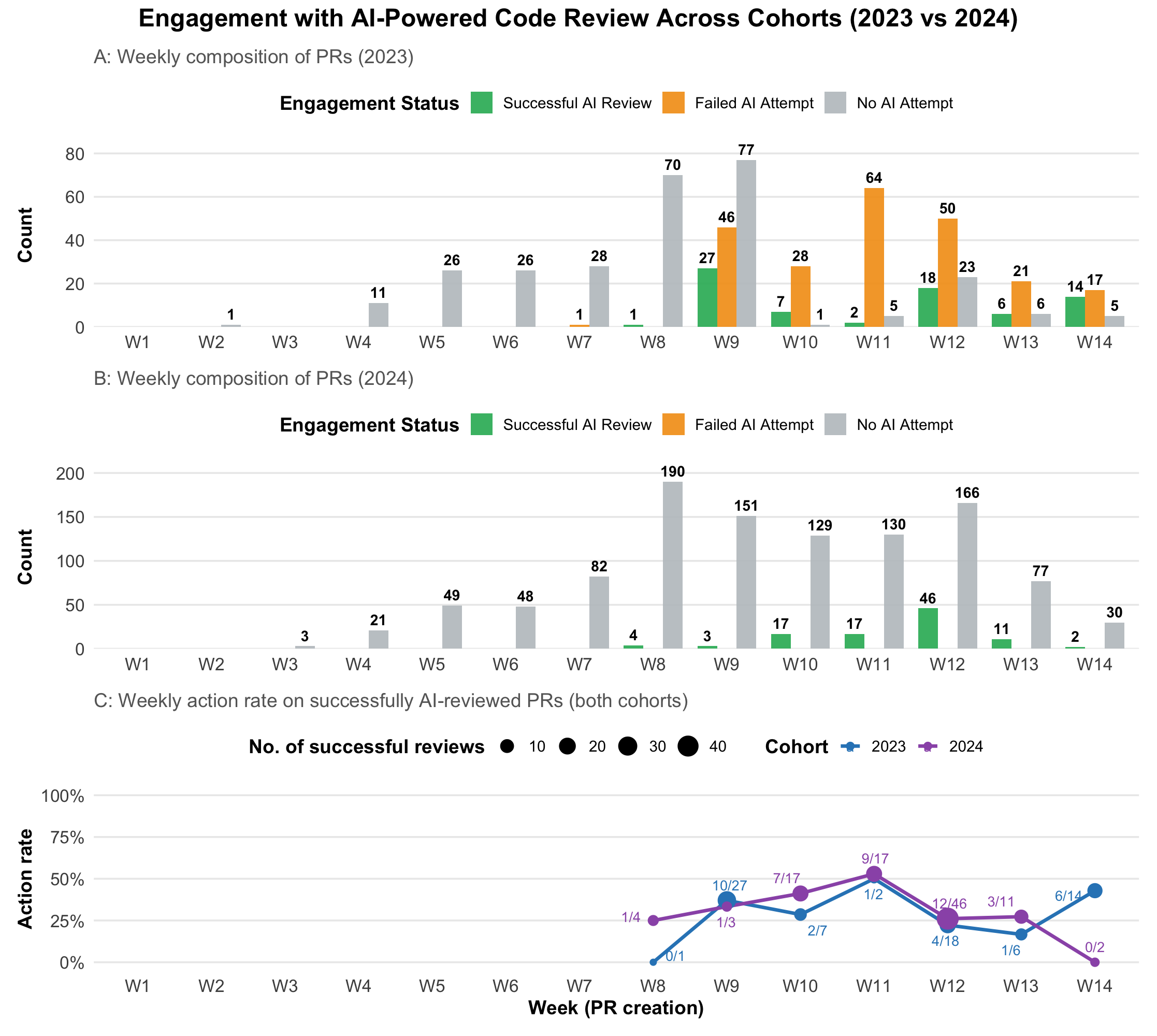} 
    \caption{A comparison of student engagement with the LLM-Reviewer tool across the 2023 and 2024 cohorts, grouped by the week of PR creation. 
    \textbf{Panels (A) and (B)} show the weekly composition of all PRs for the 2023 and 2024 cohorts, respectively, categorised by engagement status: Successful AI Review (green), Failed AI Attempt (orange), or No AI Attempt (grey). 
    \textbf{Panel (C)} plots the weekly Action Rate for both cohorts, representing the percentage of successfully reviewed PRs that were followed by at least one new code commit. The size of each data point in Panel (C) is proportional to the number of successful AI reviews in that week for each cohort.}
    \label{fig:engagement}
\end{figure*}

\smallsection{Result}
Table~\ref{tab:rq1-summary-by-cohort} summarises engagement across cohorts and points to two macro trends. First, overall development activity and process maturity increased markedly in the second year: the 2024 cohort produced more than twice the PRs (1176 vs.\ 581) and substantially more commits and comments, reflecting stronger iterative, agile practices. Second, the technical friction evident in 2023 was effectively removed in 2024: failed AI attempts fell from 227 to zero.

We treat a \emph{failed AI attempt} as an operational marker of technical friction: a bot-triggered review that yields no usable feedback (e.g., invalid/unauthorised API credentials, submitting non-code artefacts, exceeding file limits, or over-broad “whole-repo” invocations). Such failures interrupt the PR-centric feedback loop and delay monitoring/evaluation within the SRL cycle. Between 2023 and 2024 we mitigated this friction through targeted changes: pre-checks and scope guardrails (to block pathological requests), clear and actionable error messages (how to fix secrets, acceptable file types/paths, size limits), demonstrated examples for “right-way” to conduct AI-assisted code reviews in our lecturers, and highlighted explicit human-in-the-loop framing during lectures and in the GitHub READMEs. By 2024, these changes eliminated measured failures without adding process overhead.

Adoption patterns also shifted. While most teams used the tool in 2023 (93\%), about half did so in 2024 (50\%). Two factors likely contributed. First, novelty effects in early 2023 encouraged broad experimentation with an LLM reviewer; by 2024, students were more familiar with AI (e.g., in IDEs like Copilot) and used our reviewer more selectively. Second, the much higher PR volume in 2024 suggests faster cadence and more strategic use of AI review on PRs perceived as higher risk or higher impact, rather than uniformly across all PRs.

\begin{table}[t]
  \centering
  \caption{Summary of RQ1 engagement metrics by cohort (2023 vs 2024).}
  \label{tab:rq1-summary-by-cohort}
  \begin{tabular}{lrr}
    \toprule
    \textbf{Metric} & \textbf{2023} & \textbf{2024} \\
    \midrule
    Participating Teams                    & 29  & 34   \\
    Teams Using AI Tool                    & 27  & 17   \\
    Total Pull Requests (PRs)              & 581 & 1176 \\
    PRs with Successful AI Review          & 75  & 100  \\
    PRs with Failed AI Attempt             & 227 & 0    \\
    Actioned PRs (after successful review) & 24  & 33   \\
    Overall Action Rate (\%)               & 32  & 33   \\
    Total Commits in PRs                   & 8699 & 9436 \\
    Total Comments in PRs                  & 1698 & 2872 \\
    \bottomrule
  \end{tabular}
\end{table}

Figure~\ref{fig:engagement} Panels~A and B show the weekly composition of PRs by engagement status. PR creation in both cohorts follows the course’s agile cadence: a gradual build-up through Weeks~3–7, and a sharp increase in Weeks~8-12. AI usage can officially begin in Week~7, after our instructional push for code quality and code reviews. In both years, we see increases in AI usage after Week 8 (the end of Sprint 2 development) due to students having more time to concentrate in this newly introduced activity. A key cross-sectional finding is the significant increase in overall activity in the second year; the 2024 cohort generated substantially more PRs (e.g., 190 in Week 8) compared to the 2023 cohort (70 in Week 8), indicating a higher degree of iterative development.

A notable 2023 episode was an over-reliance spike: once repositories were correctly connected, several teams attempted “review everything” submissions in Week~11, inflating failures and briefly disrupting SRL. Several teams attempted to submit nearly entire repositories for inspection in a single run, which exceeded limits and inflated the Week~11 failures. We responded by tightening guardrails and returning formative, targeted messages that coached students to scope meaningful diffs, not entire repositories. Usage stabilised in Weeks~13–14 with smaller, successfully reviewed PRs and fewer failures, evidence that timely, actionable system feedback helps sustain student agency and prevents cognitive offloading.

Perhaps the most significant finding relates not to the frequency of use, but to its impact on student actions. Panel C of Figure \ref{fig:engagement} shows the Action Rate, our proxy for student responsiveness. In this paper, the action rate is computed weekly as: among PRs that received a successful AI comment in week \textit{W} (by PR creation week), the proportion that had at least one subsequent commit to the same PR after the timestamp of the first AI comment. Despite the differences in adoption and activity between the cohorts, the pattern of responsiveness was remarkably consistent. Both the 2023 (blue) and 2024 (purple) cohorts exhibited a nearly identical trend: (i) an initial rise in responsiveness, peaking around Weeks 9-11 as teams incorporated feedback during active development, and (ii) a noticeable dip in Week 12, coinciding with the end of the final major development sprint when teams were likely focused on integration and finalising features. However, a notable divergence occurs in Week 14. The 2023 cohort's action rate increased sharply to 43\%, suggesting a final, concentrated push to address code quality before project handover. We believe this reflects the development patterns of that initial cohort, which conducted fewer PRs overall (Panel A) and may have been less invested in continuous code quality from the start. In contrast, the action rate for the 2024 cohort dropped to zero in the final week. This difference is likely explained by an evolution in our pedagogical focus. In 2024, a greater emphasis was placed on continuous quality control throughout the project. This is reflected in the significantly higher PR volume from the 2024 cohort (Panel B). We hypothesise that this more consistent and proactive approach to quality assurance meant that improvements were integrated steadily, eliminating the pressure and the need for a last-minute "quality crunch" in the final week.

\begin{table*}[t]
\footnotesize
\setlength{\tabcolsep}{3pt}
\renewcommand{\arraystretch}{1.1}
\caption{Team-AA, PR \#129 (2024): AI feedback and subsequent student actions (UTC).}
\label{tab:si-koala-129}
\begin{tabularx}{\linewidth}{@{}l l l l X@{}}
\toprule
\textbf{Event} & \textbf{Timestamp} & \textbf{Actor} & \textbf{Comments (condensed)} \\
\midrule
Bot review/comment & 2024-05-14 07:08:35 & github-actions[bot] &
“ChatGPT review for \texttt{.../Form.js}: 1) Documentation defects: comments are sporadic and not detailed; …” \\
Bot review/comment & 2024-05-14 07:08:36 & github-actions[bot] &
“ChatGPT review for \texttt{.../ManuscriptSubmission.js}: 1) Documentation defects: lack of JSDoc comments; …” \\
Bot review/comment & 2024-05-14 07:08:36 & github-actions[bot] &
“ChatGPT review for \texttt{.../ThirdPageSubmission.js}: 1) Documentation defects: limited documentation; …” \\
Bot review/comment & 2024-05-14 07:08:37 & github-actions[bot] &
“ChatGPT review for \texttt{.../submission.service.js}: add inline/JSDoc comments; clarify function purposes; …” \\
\midrule
Follow-up commit & 2024-05-14 07:14:21 & student team &
Frontend styling/refactoring; fix errors/warnings; improve form handling. \\
Follow-up commit & 2024-05-14 07:20:07 & student team &
Add Axios interceptor; propagate user info to token manager; set request headers. \\
Follow-up commit & 2024-05-14 07:24:38 & student team &
Fix login email domain validation. \\
Follow-up commit & 2024-05-14 07:28:26 & student team &
Improve form handling and readability across components/utils. \\
Follow-up commit & 2024-05-14 07:31:57 & student team &
Fix ESLint errors and warnings. \\
Follow-up commit & 2024-05-14 07:34:44 & student team &
Fix \texttt{useFormSubmit}; improve form handling. \\
Follow-up commit & 2024-05-14 07:37:03 & student team &
Add sign-in failure message. \\
Follow-up commit & 2024-05-14 07:39:11 & student team &
Refresh-token integration. \\
Follow-up commit & 2024-05-14 07:42:33 & student team &
Form error handling; housekeeping. \\
Follow-up commit & 2024-05-14 07:47:11 & student team &
\texttt{fix(lint)}: address issues; simplify code. \\
\bottomrule
\end{tabularx}
\end{table*}

Quantitatively, the overall responsiveness was also highly consistent. In 2023, 32\% of successfully reviewed PRs were actioned (24 out of 75). In 2024, this figure was nearly identical at 33\% (33 out of 100). 
To ground the action-rate metric in a concrete instance of SRL-in-action, Table~\ref{tab:si-koala-129} presents a fine-grained trace from \emph{Team-AA}, PR~\#129 (2024). A PR is marked \emph{actioned} if it has at least one commit with a timestamp strictly after the first successful AI comment on that PR. In PR~\#129, the bot posted four file-scoped reviews at 07:08 UTC highlighting missing or sparse documentation and readability concerns across the frontend and backend. Within six minutes, students began a concentrated burst of ten commits over \(\sim\)33 minutes (07:14–07:47 UTC), addressing code refactoring and styling, ESLint fixes, form handling, token headers/auth, and error handling—changes that directly align with the AI’s feedback categories (documentation/readability, structure, and checks). Rather than trying to measure SRL, we report observable proxies: AI feedback followed by targeted commits on the same PR. This pattern accords with monitor/evaluate → plan → act practices and offers a concrete counterpart to the weekly action-rate aggregates in Figure~\ref{fig:engagement}.

\subsection{RQ2: Student perceptions of AI utility and learning}
\smallsection{Approach}
To address RQ2, we performed an in-depth analysis of students' self-reported experiences from the 2023 cohort. We used two complementary data sources: mandatory individual reflection reports, which provided rich qualitative insights, and a voluntary survey, which offered supporting quantitative data.
Notably, this perceptual analysis is confined to the 2023 cohort. This was a deliberate pedagogical decision made in response to direct student feedback. In the university's end-of-semester subject experience surveys for 2023, students commented on the high assessment workload for our subject. Acknowledging this and acting on their feedback, we made a pedagogical decision to remove the individual reflection report task in the 2024 (alongside with other changes in our sprint-based rubrics), offering to reduce the assessment burden on students. Therefore, while our behavioural analysis in RQ1 is cross-sectional, our perceptual analysis for RQ2 focuses on the rich dataset collected during the initial year of the intervention.

Our primary data source consisted of individual reflection reports. As this task was worth 8\% of the total subject grade in 2023, we achieved a 100\% submission rate from all participants in 2023 (80), providing a complete and detailed qualitative dataset. These reports captured students’ experiences across the entire code review process. This was supplemented by a voluntary survey completed by 16 students from the same cohort, which provided quantitative context for their perceptions.
When presenting the results, we foreground the qualitative insights from the reflection reports, supplementing them with survey findings where relevant.

\begin{table*}[htbp]
\caption{A summary of themes, sub-themes, and descriptions derived from the thematic analysis of survey results.}
\resizebox{\textwidth}{!}{
\begin{tabular}{|l|l|l|}
\hline
\multicolumn{1}{|c|}{\textbf{Theme}}                                  & \multicolumn{1}{c|}{\textbf{Sub-theme}}            & \multicolumn{1}{c|}{\textbf{Brief Description}}               \\ \hline
\multirow{4}{*}{\textbf{T1: Learning and Skill Development}}          & T1.1: Language Acquisition and Code Comprehension  & Understanding unfamiliar syntax, packages, and legacy code.   \\ \cline{2-3} 
                                                                      & T1.2: Security-Aware Coding Practices              & Learning to protect secrets and avoid hard-coded credentials. \\ \cline{2-3} 
                                                                      & T1.3: Alternative Problem-Solving Approaches       & Exploring new ways to approach and implement solutions.       \\ \cline{2-3} 
                                                                      & T1.4: Support for Novice and Intermediate Learners & Reinforcing learning while encouraging independent thinking.  \\ \hline
\multirow{3}{*}{\textbf{T2: Code Quality and Documentation}}          & T2.1: Stylistic Improvements and Code Hygiene      & Enhancing formatting, naming conventions, and code structure. \\ \cline{2-3} 
                                                                      & T2.2: Documentation Support                        & Generating or refining explanatory notes and summaries.       \\ \cline{2-3} 
                                                                      & T2.3: Actionable Review Output                     & Receiving clear, organized suggestions for improvement.       \\ \hline
\multirow{4}{*}{\textbf{T3: Accelerated Debugging and Understanding}} & T3.1: Time-Efficient Bug Diagnosis                 & Quickly identifying and resolving bugs.                       \\ \cline{2-3} 
                                                                      & T3.2: Explanation of Incorrect Behavior            & Understanding why code fails and how to fix it.               \\ \cline{2-3} 
                                                                      & T3.3: Clarification of Complex Code                & Parsing and explaining intricate or legacy functions.         \\ \cline{2-3} 
                                                                      & T3.4: Multi-Perspective Evaluation                 & Reviewing code from multiple technical angles.                \\ \hline
\multirow{5}{*}{\textbf{T4: Limitations}}                             & T4.1: Limited Context Awareness                    & Difficulty evaluating logic without full system context.      \\ \cline{2-3} 
                                                                      & T4.2: Technical Constraints                        & Word and response limits restrict review depth.               \\ \cline{2-3} 
                                                                      & T4.3: Contradictory Feedback                       & Occasional inconsistencies in suggestions.                    \\ \cline{2-3} 
                                                                      & T4.4: Over-Reliance Concerns                       & Avoiding dependence on AI for critical decisions.             \\ \cline{2-3} 
                                                                      & T4.5: Privacy and Cost Risks                       & Hesitation to submit full codebases due to exposure or cost.  \\ \hline
\end{tabular}
}
\label{tab:survey_results_1}
\end{table*}

\begin{figure*}[t]
    \centering
    \includegraphics[width=0.85\textwidth]{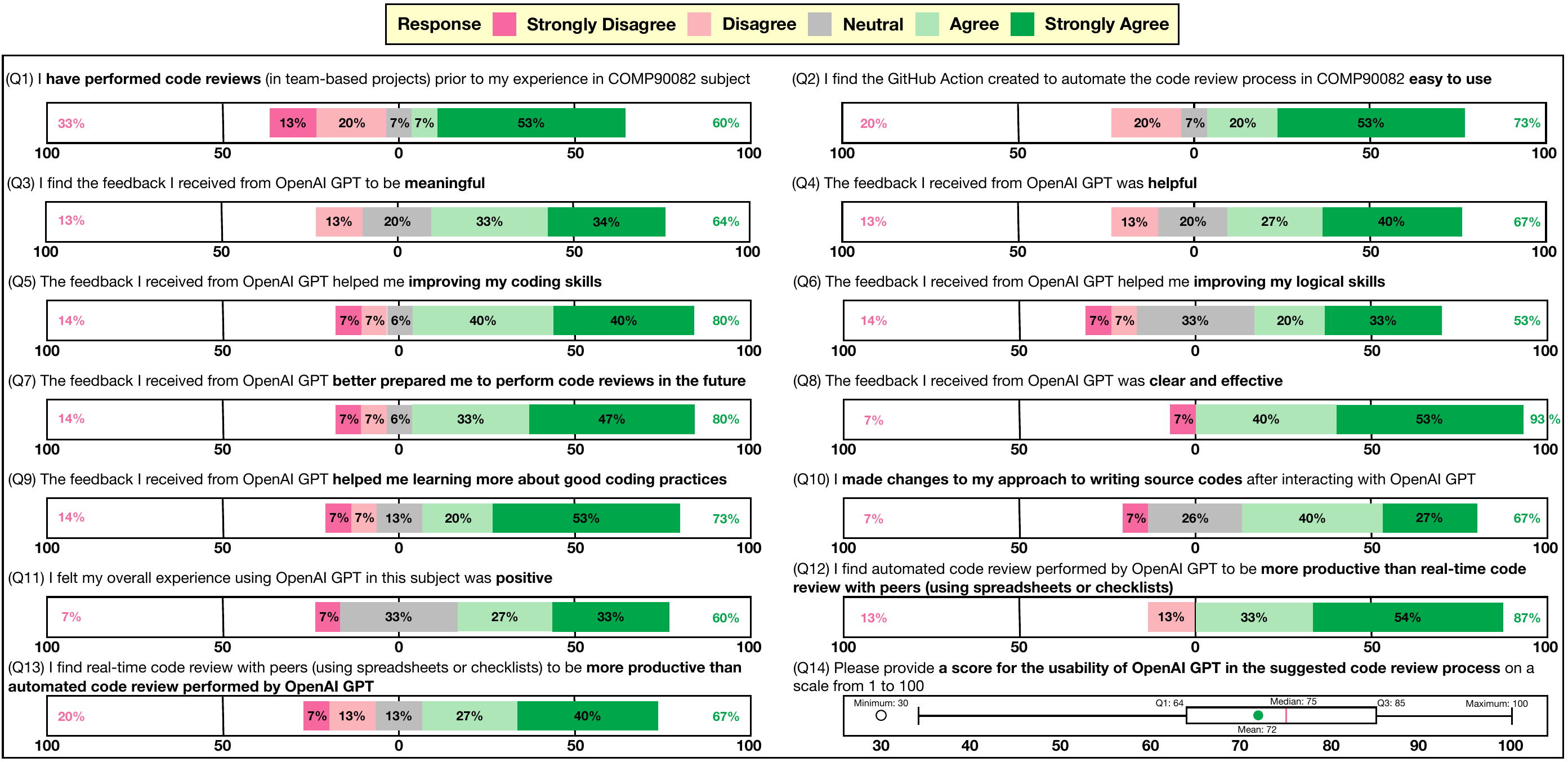}
    \caption{A summary of the survey questions and the results.}
    \label{fig:survey_results_1}
\end{figure*}

\smallsection{Result}
Through our thematic analysis of students’ reflection reports, we identified four major themes: Learning and Skill Development, Code Quality and Documentation, Accelerated Debugging and Understanding, and Limitations. These are summarised in Table~\ref{tab:survey_results_1} and form the core of our findings. We present these qualitative insights below, supplemented by complementary quantitative results from the survey (Figure~\ref{fig:survey_results_1}).

\textbf{AI as a catalyst for learning, not a replacement for thinking:} Students’ reflection reports consistently described our LLM reviewer as a learning scaffold. Across projects and languages, they used the tool as a review companion to accelerate language acquisition, decode unfamiliar packages or legacy code (\textbf{T1.1}: \textit{``AI-powered code review helps us understand unfamiliar sections from the previous team by including those files in our GitHub Actions workflow, which helps us quickly grasp the code's meaning and functionality.''}), practice security-aware habits (\textbf{T1.2}: \textit{``I mostly gained knowledge with learning good practices such as creating a \texttt{.env} file to hide open secret keys as opposed to hard coding them into files.''}), and explore alternative approaches to problems (\textbf{T1.3}: \textit{``With each interaction, I became better at coming up with new ways to approach problems and how I could code up these solutions, which has led to a noticeable improvement in my overall coding proficiency.''}). 

These qualitative insights are complemented by survey findings, which indicate strong student endorsement of the AI-assisted code review tool as a catalyst for learning: \textbf{80\%} agreed it helped improve their coding skills and better prepared them to perform code reviews in the future \textbf{(Q7)}, while \textbf{73\%} reported that it enhanced their understanding of good coding practices \textbf{(Q9)}.

\textbf{Taken together, these reflections suggest that students viewed the tool as especially valuable for novices and intermediates, where it could reinforce learning while still encouraging independent thinking} (\textbf{T1.4}: \textit{``From the code review which I did, I can definitely say that it is best to use ChatGPT if you are a beginner or intermediate-level person in coding because it helps you to simultaneously help and think about the code you are writing.''}).

\textbf{Code quality and documentation:}
Students’ reflection reports highlighted that the AI-assisted code review process led to tangible improvements in code quality and documentation. Stylistic feedback was especially valued, with students noting gains in formatting and naming conventions (\textbf{T2.1, T2.2}): \textit{``The feedback was in detail and very meaningful, giving new insights about the areas where I need to improve my code style, follow good indentation practices, give examples of how a project structure should be maintained, and naming conventions for variables, function names, widgets, etc.''}  

Beyond style, students used the tool to generate or refine documentation and explanatory notes: \textit{``ChatGPT not only provides a summary of the given code's functionality but also offers helpful suggestions or alternative code changes.''} Review output was consistently described as clear and actionable (\textbf{T2.3}): \textit{``For code reviews, ChatGPT can quickly provide a clear and organised report with suggestions for improvement.''}  

These qualitative insights are complemented by survey responses, which highlighted the tool’s usability and impact: \textbf{73\%} of students agreed it was easy to use \textbf{(Q2)}, \textbf{64\%} found the feedback meaningful \textbf{(Q3)}, and \textbf{67\%} considered it helpful \textbf{(Q4)}.

In summary, these results suggest that students not only improved their technical writing and code hygiene but also developed practical habits for structuring, documenting, and reviewing code. The reflections indicate that students often treated the AI feedback as a prompt to revisit and refine their own work, which fostered greater awareness of coding conventions and documentation practices. \textbf{In this way, the tool acted as a scaffold that supported self‑regulated learning, helping students to reflect on their process and gradually internalise higher standards of code quality.}

\textbf{Accelerated learning by lowering barriers to understanding:}
Students’ reflection reports highlighted the tool’s role in saving time and clarifying complex code. Several students emphasised efficiency gains and clearer explanations (\textbf{T3.1, T3.2}): \textit{``ChatGPT shortened this process significantly by providing explanations for why the incorrect behaviour occurred and solutions for rectifying the mistakes.''}  

Others noted its ability to parse and explain intricate functions (\textbf{T3.3}): \textit{``Another benefit of ChatGPT is that it can understand what complicated functions are doing very quickly. I wasn’t exactly sure what a function within the file did, so I asked ChatGPT, which thoroughly explained the code.''} Students also described the reviews as comprehensive and multi‑faceted (\textbf{T3.4}): \textit{``ChatGPT provides a very comprehensive check of code operation from multiple perspectives, including code writing, structure, functionality, memory management, etc.''}  

These qualitative insights are complemented by survey responses, with \textbf{87\%} of students agreeing that the AI‑powered code review process was more productive than manual checklist‑based review practices \textbf{(Q12)}, and \textbf{93\%} finding the feedback clear and effective \textbf{(Q8)}.

In short, these results show that the tool helped students identify bugs, explain incorrect behaviour, and understand complex code—making the debugging cycle faster. \textbf{By prompting students to interpret explanations and apply fixes, the process also engaged elements of self‑regulated learning, particularly reflection on errors and adjustment of problem‑solving strategies.}

\textbf{Limitations:}  
Students’ reflection reports surfaced several constraints that moderated the value of our LLM reviewer, including issues of correctness, interaction limits, dependency concerns, and privacy risks. The most salient theme, however, was the tool's lack of contextual awareness. 
Some questioned the tool’s ability to reason about system‑level logic or provide consistently accurate suggestions when the full application context was missing (\textbf{T4.1}): \textit{“For logic error, I am not sure ChatGPT had the context of the whole application to perform evaluation of logic.”} and \textit{“Without the full context of the system architecture, ChatGPT can recommend incorrect fixes to bugs in your code.”}
Students also pointed to technical constraints that restrict processing of large or interconnected codebases (\textbf{T4.2}): \textit{“ChatGPT has a word and response limit.”}
Even with added context, its capacity to evaluate logic across files or modules remained limited. One student described a contradictory interaction (\textbf{T4.3}): \textit{“ChatGPT suggested removing duplicated code, but when we asked where the duplication was, it contradicted itself and said there was none.”}
Concerns about dependence and privacy were noted as well; some students consciously avoided over‑reliance (\textbf{T4.4}): \textit{“We made a conscious effort to avoid creating overly strong dependencies on OpenAI, and just fill in gaps and improve myself through them.”} Others raised risks related to exposing proprietary code and potential analysis costs (\textbf{T4.5}): \textit{“Although I did not submit my whole codebase to ChatGPT, there could be strong negative effects to doing so, such as exposing proprietary code and the cost associated with ChatGPT analysing potentially hundreds of files.”}

These qualitative insights are complemented by survey responses. \textbf{53\%} of students agreed that the tool’s feedback helped improve their logical reasoning skills \textbf{(Q6)}, aligning with reflections about limits in logic evaluation and contextual accuracy. Nevertheless, 60\% positive ratings and a median usability of 75 show the tool is still useful when students stay in control, frame smaller diffs, and use AI comments to focus discussion. The implication is to teach students when and how to invoke the AI-reviewer, and to pair AI feedback with human dialogue for validation \textbf{(Q14)}.


\section{Lessons Learned}
\label{sec:lessons}

We synthesise insights from our RQ into 5 broader lessons that generalise across cohorts (2023, 2024) and speak directly to gaps identified in \autoref{sec:background}: (i) the need to connect AI feedback to SRL processes rather than tool use per si, (ii) the scarcity of objective behavioural evidence beyond perceptions, and (iii) operational frictions and over-reliance patterns that existing studies don't often quantify.

\textbf{1. Align AI review with agile cadence to trigger SRL}:
Engagement was punctuated by sprint peaks rather than uniform use (\autoref{fig:engagement}), and follow-up commits after successful AI comments (Panel~C) indicate evaluation $\rightarrow$ action within the SRL cycle. Embedding AI review at natural sprint gates (e.g., Definition of Done, pre-merge checks) provides time- and task-contingent cues for planning, monitoring, and evaluating, addressing the literature's call to situate AI feedback inside authentic workflows rather than as a standalone tool~\cite{bacchelli2013,oliveira2023ai}. Practically, educators should treat AI review as an established agile activity (e.g., part of PR templates and CI checks) to ensure SRL is enacted at the right granularity (diffs/files tied to user stories), not opportunistically.

\textbf{2. Reduce operational friction through tool--pedagogy co-design}:
Early high failure rates (invalid credentials, non-code artefacts, file limits) show that access alone is insufficient. Cohort-to-cohort improvements followed targeted onboarding and actionable error messages, converting friction into learnable practice. This responds to a gap in prior work that reports student perceptions but rarely quantifies deployment barriers or their remediation~\cite{baresi2025students}. We found that small design choices--clear scoping examples in PR templates; guardrails that reject oversized or mixed-content submissions with specific guidance; and visible status reporting in CI--materially change adoption curves (\autoref{fig:engagement}). Recommendation: co-design technical affordances and instruction so students learn the operational practice of code review (what to submit, when, and why), not merely how to invoke a model.

\textbf{3. Preserve student agency; prevent over-reliance with scope and guardrails}:
Once repositories were connected, some teams attempted to offload judgment by sending entire repos, reflecting the over-reliance risk noted in the literature but rarely evidenced with trace data~\cite{parra2025towards}. Constraining review scope (diff- or file-level), surfacing ``human-in-the-loop'' prompts, and giving formative error feedback recalibrated behaviour: later weeks show more targeted, purposeful reviews (\autoref{fig:engagement}). Pedagogically, this centres agency: students decide \emph{what} to inspect and \emph{how} to respond, with AI as a structured critic rather than a decider. This framing aligns with SRL's emphasis on learner control and metacognitive regulation~\cite{zimmerman2002becoming, de2025development, lopez2025dynamics, cukurova2025interplay}, countering the risk that AI displaces planning and monitoring.

\textbf{4. Measure what matters: telemetry plus structured feedback}:
Repository traces allowed us to move beyond self-report and link feedback to concrete actions. The Action Rate metric (32\% in 2023; 33\% in 2024) shows stable, semester-level responsiveness (\autoref{fig:engagement}, \autoref{tab:rq1-summary-by-cohort}), addressing the gap around tracing AI suggestions to code changes~\cite{parra2025towards}. The checklist-structured prompt (\autoref{fig:prompt}) increased explainability and made it easier to map comment categories to follow-up commits—closing the SRL loop with auditable evidence. Recommendation: pair structured AI outputs with repository-based indicators (e.g., post-feedback commits, file-level diffs) to evaluate learning processes, not just outcomes. This provides educators with a reusable, theory-aligned instrumentation strategy and avoids over-reliance on perceptions.

\textbf{5. Frame the AI as a 'companion', not an oracle, to foster critical assessment}:
Our qualitative findings from RQ2 reveal an important dynamic: students did not treat the AI tool as an infallible authority. Instead, they developed a sophisticated, critical stance on its capabilities. They found it valuable for scaffolding foundational and stylistic skills, praising its ability to improve code hygiene, clarify syntax, and accelerate debugging (\autoref{tab:survey_results_1}, Themes T2 \& T3). However, they remained highly skeptical of its ability to handle complex, system-level logic. This is evidenced by the fact that only 53\% of students agreed it helped improve their logical reasoning skills (Q6), a perception reinforced by their written reflections about the tool's lack of contextual awareness (Theme T4). This is not a failure of the tool, but a success of the learning process. It demonstrates that students, rather than becoming cognitively lazy, learned to selectively trust the AI. They successfully identified its strengths (pattern recognition, style conventions) and weaknesses (deep logic, context), and in doing so, practiced the essential professional meta-skill of knowing how to effectively collaborate with an imperfect AI. Recommendation: Educators should explicitly frame AI review tools as specialised, context-blind collaborators, not as all-knowing oracles. Set clear expectations from the outset: the tool is an expert at identifying stylistic, structural, and common implementation issues, but the final authority on complex logic and system architecture must remain with the human developer. This approach fosters healthy skepticism, prevents misplaced trust, and powerfully reinforces that the student's own critical judgment is their most important tool.

Together, these five lessons suggest a programmatic approach: (1) integrate AI review into agile ceremonies to cue SRL; (2) co-design guardrails and pedagogy to reduce friction; (3) protect agency through scope control and human-in-the-loop messaging; (4) assess impact with auditable traces rather than perceptions alone; and (5) frame the AI's role explicitly to foster critical assessment and prevent blind acceptance. This responds directly to the research gaps identified in \autoref{sec:background} and offers a replicable pathway for responsible, learning-centred deployment in capstone settings.

\subsection{Directions for Future Research}
This experience report opens several promising directions for future work. First, there is a clear need to develop more context-aware AI review tools tailored for educational settings. The limitations students noted regarding the tool's lack of system-level understanding point to a significant technical and pedagogical challenge. Future tools could be designed to ingest more of a project's context or even adapt their feedback based on the student's progress, providing a more intelligent and personalised scaffolding. Second, our study provides a baseline for cross-sectional and comparative analyses. Future research could track students over multiple semesters or years to determine whether the SRL skills fostered by AI feedback are retained and transferred to other contexts. Comparative studies could also explore the effectiveness of different AI prompting strategies. For example, would a Socratic-style AI that asks probing questions rather than providing direct suggestions lead to deeper reflection and learning? Moreover, our results have shown that the intervention supports the performance and reflection phases of SRL; the forethought phase could also be supported by modifying the system prompt to suggest new features and help the students plan the subsequent "sprint".  Finally, the complementarity we observed between human and AI review warrants further investigation. Future work could systematically study how teams blend these two modalities, examining how they negotiate feedback from different sources and which types of issues are best suited for each review method. Such studies would provide deeper insight into fostering effective, hybrid review workflows that prepare students for the collaborative realities of modern software engineering.  
\section{Threats to Validity}
\label{sec:threats}
\textbf{Threats to the external validity} relate to the single institutional context of our study. Because the work was conducted in our Master of IT capstone subject at the University of Melbourne, the findings may not generalise to other disciplines or settings. To mitigate this, our AI-powered code review GitHub Action is fully open-sourced at \url{https://github.com/agogear/chatgpt-pr-review}, enabling future studies to expand and apply the tool in diverse contexts.

\textbf{Threats to the internal validity} relate to model checkpoint variation across years. The 2023 cohort used GPT‑3.5, while the 2024 cohort used GPT‑4, introducing differences in feedback quality that could influence both engagement and perceptions. To mitigate this, we analysed engagement metrics such as adoption and action rates separately for each cohort before comparing trends, ensuring that model differences did not distort the overall patterns. For perceptions, we focused our survey and reflection analysis on the 2023 cohort only, so that all students were exposed to the same model, avoiding variation in feedback quality as a confounding factor.
\section{Conclusion}

This experience report demonstrates that AI-powered tools can be successfully integrated into software engineering education as effective scaffolds for learning. Our evidence from two semester-long cohorts shows that a deliberately designed intervention -- one that provides critiques without direct solutions -- consistently triggers an SRL cycle of evaluation and action, which also aligns well with agile methodologies where reflection and iteration are central. We found that students did not become passive recipients of feedback; instead, they learned to critically engage with the AI, augmenting their skills without sacrificing their own agency. This is a relevant finding in light of existing evidence that ---when it comes to student-AI-interaction--- high-achieving students are  more likely to engage in more frequent reflection and self-regulation \cite{Nguyen2024-jy}, and to incorporate AI’s suggestions into the key stages of task performance rather than treating them as peripheral support \cite{Kim2025-xb}. In addition to the empirical findings, the primary contribution of this work is a set of evidence-based lessons showing that the pedagogical value of an AI tool lies not in the technology itself, but in the careful co-design of the tool, task, and environment to keep the student firmly in control of their own learning process.


\bibliographystyle{ACM-Reference-Format}
\bibliography{reference}




\end{document}